\documentclass[%
 reprint,
superscriptaddress,
 amsmath,amssymb,
 aps
]{revtex4-2}
\usepackage{float}
\usepackage{graphicx}
\usepackage{dcolumn}

\usepackage{bm}

\usepackage{caption}
\usepackage{ragged2e} 
\DeclareCaptionJustification{justified}{\justifying}

\usepackage{microtype}
\usepackage{color}
\usepackage[dvipsnames]{xcolor}

\begin{document}


\title{A validated lumped-element model for bioinspired acoustic flow sensing toward the performance limit}

\author{Wei Sun}
 \affiliation{Department of Mechanical Engineering,  State University of New York at Binghamton, Binghamton, NY 13902, United States}

\author{Wanyin Zheng}
 \affiliation{Department of Mechanical Engineering,  State University of New York at Binghamton, Binghamton, NY 13902, United States}

 \author{Xiangyu Wei}
 \affiliation{Department of Mechanical Engineering,  State University of New York at Binghamton, Binghamton, NY 13902, United States}

\author{David A. Czaplewski}%
\affiliation{Center for Nanoscale Materials, Argonne National Laboratory, Argonne, IL 60439, United States}

\author{Ronald N. Miles}
\affiliation{Department of Mechanical Engineering,  State University of New York at Binghamton, Binghamton, NY 13902, United States}

\author{Jian Zhou}
\email{jianzhou@binghamton.edu}
\affiliation{Department of Mechanical Engineering,  State University of New York at Binghamton, Binghamton, NY 13902, United States}

\date{\today}

\begin{abstract}
Flow sensing is fundamental to both biological survival and technological innovation. Inspired by biological mechanoreceptors, artificial flow sensors detect subtle fluid motion using slender, viscous-driven structures. Among these, acoustic flow sensors that mimic nature’s velocity-sensitive ears have the potential to transform vector sound detection. 
Yet, despite their potential, understanding of how design parameters determine ultimate sensor performance remains limited. To effectively guide flow sensor design, we develop and experimentally validate a lumped-element model that captures the broadband motion of slender microcantilevers immersed in fluid, combining analytical simplicity with quantitative accuracy. The model predicts flow-induced motion, thermomechanical noise, and the minimum detectable signal level, showing strong agreement with experimental measurements in air over a broad frequency range from 100~Hz to 10,000~Hz. This validated model provides a straightforward theoretical framework for designing high-performance micro- and nanomechanical sensors for flow and vector sound detection. 
\end{abstract}

\maketitle

\section{Introduction}
Flow sensing, the ability to detect and quantify motion of a surrounding medium, is fundamental to both biological survival and technological innovation. Over millions of years, evolution has produced ultrasensitive mechanoreceptors across diverse species, including fish~\cite{kindig2023asymmetric,coombs2001smart}, insects such as mosquitoes~\cite{menda2019long,pantoja2023tiny,nakata2020aerodynamic,gopfert2000nanometre}, and arachnids such as spiders~\cite{shamble2016airborne,zhou2022}. These receptors typically leverage slender, hair-like structures that interact with surrounding fluid motion, enabling functions ranging from motion control to prey localization. When tuned to minute, fluctuating flows, they act as velocity-sensitive ears that detect acoustic particle velocity, in contrast to the pressure-sensitive ears of mammals. Orb-weaving spiders, for example, use their wispy webs as auditory sensors to detect and localize distant sounds with near-optimal mechanical sensitivity~\cite{zhou2022,Zhou2017}, whereas male mosquitoes detect female wingbeats from meters away using their fluffy antennae, achieving nanometer-scale displacement resolution corresponding to particle velocities of only a few micrometers per second~\cite{menda2019long,Gopfert2000,pantoja2023tiny}.

Inspired by biological exemplars, a wide range of micro- and nano-fabricated flow sensors have been developed over the past few decades, employing diverse geometries, materials, and transduction mechanisms~\cite{tao2012hair,droogendijk2014performance,kottapalli2016nanofibril,jin2025flexible,hu2025high,li2025fish}. These compact, low-power devices can detect flow in confined or complex environments, enabling applications such as flow measurement, motion control, navigation, and environmental monitoring. More recently, bioinspired flow sensing has been introduced as a means of sound detection~\cite{Zhou2017,zhou2018highly}, guided by discoveries in spider and mosquito hearing~\cite{zhou2022,menda2019long,pantoja2023tiny}. This approach enables a new paradigm for detecting the vector nature of sound rather than its pressure. In certain performance aspects, man-made sensors have surpassed their biological counterparts. For instance, broadband acoustic flow sensing from 1~Hz to 50~kHz has been demonstrated, achieving nearly full fidelity and maximum mechanical responsivity~\cite{Zhou2017,zhou2018highly}. Despite this progress, the absolute sensitivity of man-made flow sensors~\cite{tao2012hair,droogendijk2014performance,Zhou2017,jin2025flexible,hu2025high,li2025fish}—that is, the minimum detectable flow—remains below that of biological systems such as the mosquito~\cite{menda2019long}.

Accurate and physically intuitive modeling is essential for understanding biological flow sensing and advancing engineered devices toward their physical limits. Existing modeling efforts range from simple single-degree-of-freedom approximations~\cite{tao2012hair,shimozawa1984aerodynamics,droogendijk2014performance} to detailed continuum analyses of microstructures~\cite{miles2018sound,Lai2024}. While single-degree-of-freedom formulations are intuitive and computationally efficient, they lose accuracy when applied to complex structures or broad frequency ranges. In contrast, continuum models capture distributed dynamics more comprehensively but are analytically demanding and not straightforward to interpret intuitively. Bridging this gap requires a physically transparent yet quantitatively accurate model that unifies analytical tractability with predictive power.

\begin{figure*}[htp]
\centering 	\includegraphics[width=0.94\textwidth]{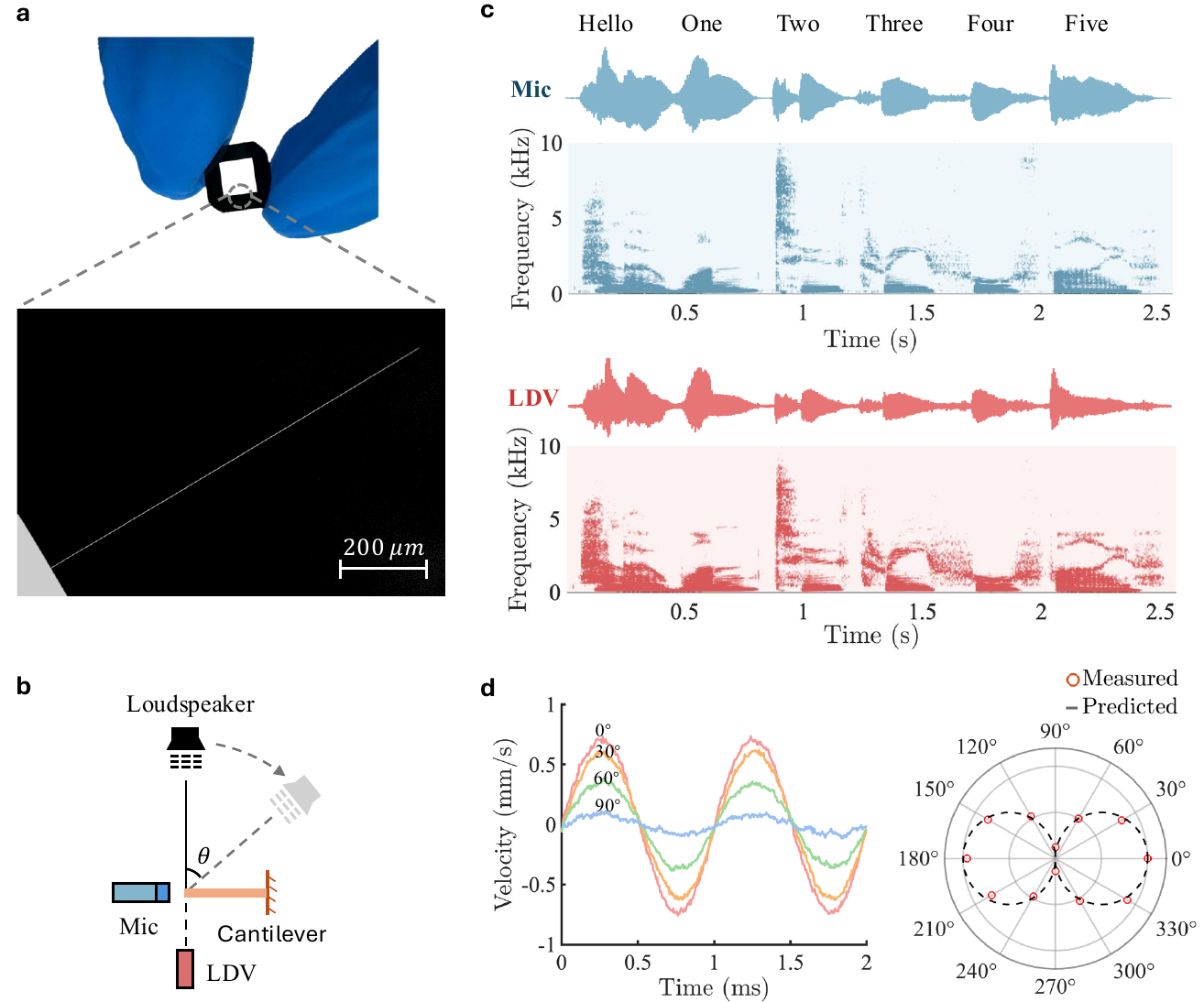}
\caption{\textbf{Acoustic flow sensing with slender microcantilevers demonstrating high fidelity and intrinsic directionality.} \textbf{a}, Silicon nitride microcantilevers fabricated using a mass-producible semiconductor process. The top optical micrograph shows a fabricated chip (10~mm~$\times$~10~mm) with microcantilevers arranged within a central opening (5~mm~$\times$~5~mm). The bottom scanning electron micrograph shows a slender microcantilever ($L = 1000~\mu$m, $b = 2~\mu$m, $h = 780~\text{nm}$). 
\textbf{b}, Schematic of the experimental setup. The sound field was generated by a loudspeaker in an anechoic chamber. The sound pressure near the chip was measured using a calibrated pressure microphone (Mic), while the tip motion of a microcantilever was measured using a laser Doppler vibrometer (LDV). 
\textbf{c}, Time-domain waveforms and corresponding spectrograms in response to a speech signal, measured by the microphone (top) and the laser Doppler vibrometer (bottom, at the microcantilever tip shown in \textbf{a}). A recorded speech signal (“Hello, One, Two, Three, Four, Five”) was reproduced by a loudspeaker positioned 3~m from the chip, with sound propagating normal to the chip surface ($\theta = 0^{\circ}$). \textbf{d}, Directional response of the microcantilever to a 1000~Hz tone at 67~dB SPL, generated by a loudspeaker positioned 0.5~m from the chip at various sound incidence angles ($\theta$).}
\label{fig.1}
\end{figure*}

Here, to efficiently guide the design of flow sensors toward their ultimate performance limits, we develop and experimentally validate a lumped-element model, building on insights from our previous continuous-system formulation~\cite{Lai2024}, that captures the broadband motion of slender microcantilevers immersed in fluid with enhanced simplicity and predictive accuracy. The model accurately predicts flow-induced motion, thermomechanical noise, and the minimum detectable signal level, showing excellent agreement with experimental measurements in air over a broad frequency range from 100~Hz to 10,000~Hz. This validated model provides straightforward theoretical guidance for the design of high-performance micro- and nanomechanical flow sensors that operate near the thermomechanical noise limit with exceptional fidelity and sensitivity.

\section{Results and Discussion}
\subsection{Bioinspired acoustic flow sensing with slender microcantilevers} 
\label{Sec:microcantilever}

Before introducing the analytical model, we first show that slender microcantilevers can detect acoustic particle velocity with high fidelity and intrinsic directionality (Fig.~\ref{fig.1}). We fabricated 780~nm-thick silicon nitride microcantilevers with lengths ranging from 500 to 1000~$\mu$m and widths from 2 to 8~$\mu$m using a mass-producible semiconductor fabrication process. The detailed fabrication procedure is provided in the Methods section. As shown in Fig.~\ref{fig.1}a, the cantilevers are arranged within a central opening of the fabricated chip. The experimental setup used to characterize the microcantilevers is illustrated in Fig.~\ref{fig.1}(b). A sound field was generated by loudspeakers, while the sound pressure near the chip was monitored using a calibrated pressure microphone (B\&K~4138). The sound-induced motion at the tip of each microcantilever was measured with a laser Doppler vibrometer (Polytec~OFV-534). All experiments were performed in an anechoic chamber (5.4~m~$\times$~4.2~m~$\times$~3.5~m). Further details of the experimental setup are described in previous studies~\cite{Zhou2017,Lai2024}. 

To assess acoustic sensing fidelity, we measured the response of slender microcantilevers (Fig.~\ref{fig.1}a) to a recorded speech signal (“Hello, One, Two, Three, Four, Five”). The speech signal was produced by a loudspeaker positioned 3~m from the chip, propagating normal to the chip surface ($\theta = 0^{\circ}$). The time-domain signals and corresponding spectrograms measured on a 1000~$\mu$m-long slender microcantilever ($b=2~\mu$m, $h=780~\text{nm}$) are shown in Fig.~\ref{fig.1}c, where the top and bottom traces correspond to the microphone and laser Doppler vibrometer measurements, respectively. The two datasets exhibit good agreement in both waveform shape and spectral content, indicating that the microcantilever captures sound with high fidelity.

Unlike sound pressure, which is a scalar quantity, acoustic particle velocity is a vector; thus, flow sensing inherently provides directional sensitivity. For a slender microcantilever, as illustrated in Fig.~\ref{fig.1}b, the response is primarily governed by the component of the acoustic particle velocity normal to its length, following $v_\theta(t) = v_0(t)\cos(\theta)$~\cite{Zhou2017}, where $v_0(t)$ denotes the response for normal incidence ($\theta = 0$).  
Figure~\ref{fig.1}d shows the measured directional response to a 1000~Hz tone at 67~dB SPL, emitted from a loudspeaker 0.5~m from the chip at various incident angles. Overall, the measured directional responses agree well with the theoretical prediction, indicating that the slender microcantilever operates as a velocity-sensitive element with intrinsic directionality. Minor deviations from the ideal theoretical response are likely caused by local flow perturbations induced by the chip frame.

\subsection{Lumped-element modeling}\label{sec:analytical modeling}
To guide sensor design, we develop a lumped-element model of a slender microcantilever immersed in fluid, representing a typical viscous-driven structure. The model captures the beam’s flow-induced dynamics, thermomechanical noise, and the corresponding minimum detectable signal level. For analytical tractability, several assumptions are made. The microcantilever has a rectangular cross-section with dimensions $L$, $b$, and $h$, compatible with conventional semiconductor fabrication processes. The beam is slender ($L \gg b,\, L \gg h$) and operates in the small-deflection regime, where linear elasticity and flexural vibrations dominate. Acoustic excitation is treated as a plane wave incident along the thickness axis, normal to both the beam’s length and width. The material is assumed isotropic and elastic, and because viscous damping from the surrounding fluid is significant, internal damping is neglected. This configuration is representative of typical flow-sensing geometries and serves as the basis for modeling.


According to Euler–Bernoulli beam theory, the transverse displacement of the beam, 
$w(x,t)$, along the thickness direction can be expressed as:
\begin{equation}
EI \frac{\partial^4 w(x,t)}{\partial x^4} + \rho A \frac{\partial^2 w(x,t)}{\partial t^2} = F(x,t),
\label{eq:Euler–Bernoulli}
\end{equation}%
The first term on the left represents the bending-induced mechanical force per unit length, where $E$ is the Young’s modulus, $I=bh^3/12$ is the area moment of inertia. The second term on the left accounts for the inertial force per unit length of the beam, where $\rho$ is the material density, $A=bh$ is the beam cross-sectional area. The right term represents the external excitation force per unit length, which, for a viscous-driven beam, takes the form \cite{miles2018sound}
\begin{equation}
F(x,t) = R(\omega)[u(x,t) - \dot{w}(x,t)]
\label{eq:F(x,t)}
\end{equation}%
where $u(x,t)$ is the fluid velocity, $\dot{w}(x,t)$ is the beam velocity, and $R(\omega)$ is the beam impedance defined as $R(\omega) = C(\omega) + i\omega M(\omega)$. Here $C(\omega)$ represents the viscous damping coefficient per unit length and $M(\omega)$ denotes the co-vibrating mass of the surrounding medium. Here we have, for convenience, assumed time dependence of the form $e^{i\omega t}$, enabling the use of a frequency dependent impedance.  Further details on the viscous force are provided in Methods. 

Before solving Eq.~(\ref{eq:Euler–Bernoulli}), it is useful to qualitatively consider how the scaling of beam dimensions affects the fluid–structure interaction. 
The terms on the left-hand side of Eq.~(\ref{eq:Euler–Bernoulli}) are proportional to either $I=bh^3/12\sim l^4$ or $A=bh\sim l^2$, where $l$ represents the scale length. While it is not straightforward to calculate the dependence of the right-hand-side term, the fluid forces generally depend on the surface area of the structure. For a beam with a rectangular cross-section, the surface area per unit length is $2(b + h)\sim l$. As the cross-sectional size decreases, the left-side terms ($\sim l^4$ and $\sim l^2$) decrease more rapidly than the right-side term ($\sim l$). For a slender beam with a sufficiently small cross-section, the viscous forces on the right-hand side will dominate the mechanical forces on the left-hand side, leading to $0\approx F(x,t)$. Consequently, $\dot{w}(x,t) \approx u(x,t)$ according to Eq.~(\ref{eq:F(x,t)}), indicating that the beam captures the surrounding fluid motion up to full fidelity and maximum physical efficiency, similar to a thin strand of spider silk~\cite{Zhou2017}.

To obtain the quantitative solution of Eq.~(\ref{eq:Euler–Bernoulli}), substituting Eq.~(\ref{eq:F(x,t)}) into Eq.~(\ref{eq:Euler–Bernoulli}) leads to
\begin{equation}
EI \frac{\partial^4 w(x,t)}{\partial x^4} + (\rho A+M) \frac{\partial^2 w(x,t)}{\partial t^2} + C\frac{\partial w(x,t)}{\partial t}= f(x,t),
\label{eq:Euler–Bernoulli2}
\end{equation}
where $f(x,t)= R(\omega)u(x,t)$. By separating variables, the beam displacement can be expressed as $w(x,t) = \sum_{j=1}^{\infty} \phi_j(x) z_j(t)$, where the subscript $j$ denotes the mode number, $\phi_j(x)$ is the normalized mode shape function, and $z_j(t)$ represents the motion of the $j$-th mode. Using the orthogonality of the mode shapes~\cite{doi:https://doi.org/10.1002/9780470117866.ch11}, Eq.~(\ref{eq:Euler–Bernoulli2}) 
can be expressed in lumped-element form, yielding the motion equation for the $j$-th mode as:
\begin{equation}
    m_{\text{eff},j} \ddot{z}_j(t) + k_{\text{eff},j} z_j(t) + c_{\text{eff},j} \dot{z}_j(t) = f_{\text{eff},j}(t)
\label{eq:lumped-element}
\end{equation}
where the effective lumped parameters are:
\begin{subequations} \label{eq:eff_parameters}
\begin{align}
m_{\text{eff},j} &= (\rho A+M) \int_0^L \phi_j^2(x)\, dx  \label{eq:meff} \\
k_{\text{eff},j} &= E I \int_0^L \phi_{j,xxxx}(x)\, \phi_j(x)\, dx  \label{eq:keff} \\
c_{\text{eff},j} &= C \int_0^L \phi_j^2(x)\, dx  \label{eq:ceff} \\
f_{\text{eff},j}(t) &= \int_0^L \phi_j(x)\, f(x,t)\, dx \label{eq:feff}
\end{align}
\label{eq:equivalent coefficients}
\end{subequations}%
\indent The analytical procedure outlined above is applicable to various boundary conditions, including cantilevered and doubly clamped beams, as well as to beams with different uniform cross-sectional geometries, such as rectangular or circular shapes. To elucidate the key design parameters that govern the performance of the velocity sensor, this study focuses on a cantilever configuration with a rectangular cross-section. For a cantilever beam, the boundary conditions for Eq.~(\ref{eq:Euler–Bernoulli2}) are the clamped and free end conditions, $w(0,t) = \frac{\partial w}{\partial x}(0,t) =0$, $\frac{\partial^2 w}{\partial x^2}(L,t) = \frac{\partial^3 w}{\partial x^3}(L,t) = 0
$. The normalized mode shape function for a cantilever is given by~\cite{doi:https://doi.org/10.1002/9780470117866.ch11}
\begin{equation}
\begin{aligned}
\phi_j(x) & = \frac{1}{2}\Big[  \cos(\beta_j x) - \cosh(\beta_j x) \\
& - \frac{\cos(\beta_j L) + \cosh(\beta_j L)}{\sin(\beta_j L) + \sinh(\beta_j L)}\cdot [\sin(\beta_j x) - \sinh(\beta_j x)] \Big]
\label{eq: mode}
\end{aligned}
\end{equation}
\noindent where $\beta_j$ is a mode-dependent coefficient determined by the cantilever boundary conditions through $\cos(\beta L) \cosh(\beta L)+1 = 0$. 
By substituting the cantilever mode shape into Eq.~(\ref{eq:equivalent coefficients}), the equivalent lumped-element parameters for each mode can be evaluated as:
\begin{subequations} \label{eq:eff_parameters2}
\begin{align}
m_{\text{eff},j} & = \frac{m_t}{4} \label{eq:meff2} \\
k_{\text{eff},j} & = EI \frac{\beta_j^4 L}{4} \label{eq:keff2} \\
c_{\text{eff},j} & = \frac{c_t}{4} \label{eq:ceff2} \\
f_{\text{eff},j}(t) &= \alpha_j r_t u(t) \label{eq:feff2}
\end{align}
\label{eq:equivalent coefficients2}
\end{subequations}%
\noindent where $m_t = (\rho A + M)L$ is the total effective mass, including both the structural mass of the beam and the added mass from the co-vibrating medium, $c_t = CL$ is the total damping of the beam, $\alpha_j = \frac{1}{L} \int_0^L \phi_j(x) dx$ is a mode-dependent coefficient; $r_t = RL$ is the total impedance of the beam, and $u(x,t)$ is simplified to $u(t)$ for spatially uniform fluid velocity along the beam length. For a cantilever beam, the maximum displacement of each mode shape $\phi_j(x)$ occurs at the free end, i.e., $|\phi_j(L)| = 1$, regardless of the mode number. Since $w_j(L,t) = z_j(t)\phi_j(L)$, it follows that $|w_j(L,t)| = |z_j(t)|$, indicating that $z_j(t)$ in the lumped-element model of Eq.~(\ref{eq:lumped-element}) directly represents the physical motion of the $j$-th mode at the beam tip. 

Based on the lumped-element model, we perform three analyses: first, the beam’s response to acoustic particle velocity excitation, second, its response to thermomechanical fluctuations, and finally, the comparison of these two responses to determine the input-referred noise, which defines the minimum detectable signal limited by thermomechanical noise.
To estimate the steady-state frequency response of the beam, we assume a harmonic particle velocity excitation $u(t) = U e^{i \omega t}$ and a corresponding beam displacement response $z_j(t) = Z_j e^{i \omega t}$. Substituting these expressions into Eq.~(\ref{eq:lumped-element}) yields the transfer function $H_j(\omega)$ between the beam-tip amplitude to the acoustic particle velocity:
\begin{equation}
H_j(\omega)
= \frac{Z_j}{U}
= \frac{\dfrac{\alpha_j r_t}{m_{\text{eff},j}}}
{\omega_j^2 - \omega^2 + i \omega \dfrac{c_{\text{eff},j}}{m_{\text{eff},j}}}
\label{eq:HUZ}
\end{equation}
\noindent where $\omega_j = \sqrt{k_{\text{eff},j} / m_{\text{eff},j}}$. The total beam velocity due to acoustic excitation can then be obtained by summing the contributions from all vibration modes, $\dot{w}(x,t) = \sum_{j=1}^{\infty} \phi_j(x) \dot{z}_j(t)$.
At the cantilever tip, the velocity can be expressed as $v(L,t) = V e^{i\omega t}=\dot{w}(L,t) = i\omega \sum_{j=1}^{\infty} \phi_j(L) z_j(t)$. Substituting $z_j(t) = H_j(\omega) Ue^{i\omega t}$ into the above expression gives the transfer function relating the beam-tip velocity $V$ to the acoustic particle velocity $U$:
\begin{equation}
\frac{V}{U}
= i\omega \sum_{j=1}^{\infty} \phi_j(L) H_j(\omega) 
\label{eq:transfer function}
\end{equation}%
\indent The thermomechanical motion of the beam is estimated using the equipartition theorem, which equates the average potential energy of each mode to thermal energy:
\begin{equation}
\frac{1}{2} k_B T = \frac{1}{2} k_{\text{eff},j} \cdot \bar{z_j}^2
\label{eq:potential energy}
\end{equation}
\noindent where $k_B$ is Boltzmann’s constant, $T$ is the absolute temperature, and $\bar{z_j}^2$ is the mean square displacement of the $j$-th mode. The mean square displacement of the $j$-th mode can be obtained by 
\begin{equation}
\bar{z_j}^2 = \int_{-\infty}^{\infty} S_{zz,j}(\omega) d\omega
\label{eq:mean square displacement}
\end{equation}
\noindent where $S_{zz,j}(\omega)$ is the power spectral density of displacement of mode $j$,
\noindent Applying Eq.~(\ref{eq:HUZ}, \ref{eq:potential energy}-\ref{eq:mean square displacement}) yields the displacement power spectral density of the $j$-th mode:
\begin{equation}
S_{zz,j}(\omega)
= \frac{4k_B T c_t}{\pi m_t^2} \cdot \frac{1}
{(\omega_j^2 - \omega^2)^2 + \left( \omega \dfrac{c_t}{m_t} \right)^2}
\label{eq:Szz2}
\end{equation}
In the special case of thermal excitation, which is assumed to be spatially uncorrelated, the random responses of the modes are also uncorrelated~\cite{Lai2024}. Consequently, the total thermomechanical displacement spectral density of the beam can be expressed as a simple summation of the power spectral densities of the individual modes:
$S_{zz}(\omega) = \sum_{j=1}^{\infty} S_{zz,j}(\omega)$. Since the velocity spectral density is related by $S_{\dot{z}\dot{z}}(\omega) = \omega^2 S_{zz}(\omega)$,
the total velocity noise spectrum can be expressed as
\begin{equation}
S_{\dot{z}\dot{z}}(\omega) = \omega^2 \sum_{j=1}^{\infty} \frac{4k_B T c_t}{\pi m_t^2}
\cdot \frac{1}
{(\omega_j^2 - \omega^2)^2 + \left( \omega \dfrac{c_t}{m_t} \right)^2}
\label{eq:Szz_dot}
\end{equation}%
\indent Lastly, the velocity-referred thermomechanical noise $S_{vv_{\text{th}}}$, can be estimated as
\begin{equation}
S_{vv_{\text{th}}} = \frac{S_{\dot{z}\dot{z}}}{|V/U|^2}
    \label{eq:Spp}
\end{equation}%

To compare the noise performance of velocity and pressure sensors, the velocity-input noise can be converted to pressure-input noise under the plane-wave assumption, where acoustic pressure and particle velocity are related by $U = P / \rho_0 c$, with $\rho_0$ denoting the air density and $c$ the speed of sound~\cite{Zhou2017}. Accordingly, the pressure-referred thermomechanical noise can be expressed as $S_{pp_{\text{th}}} = (\rho_0 c)^2 S_{\dot{z}\dot{z}} / |V/U|^2$. Substituting Eqs.~(\ref{eq:transfer function}) and (\ref{eq:Szz_dot}) into this relation yields
\begin{equation}
S_{pp_{\text{th}}}(\omega) 
= (\rho_0 c)^2 \frac{k_B T c_t}{4 \pi |r_t|^2 } 
\cdot 
\frac{\sum_{j=1}^{\infty} \dfrac{1}{\alpha_j^2} 
| H_j(\omega)|^2}{
| \sum_{j=1}^{\infty} \phi_j(L) H_j(\omega)|^2}.
\label{eq:Spp3}
\end{equation}

For each vibration mode $j$, Eq.(\ref{eq:Spp3}) can be simplified as  
\begin{equation}
    S_{pp_{th},j} 
= (\rho_0 c)^2\frac{k_B T c_t}{4 \alpha_j^2 \pi |r_t|^2}
\label{eq:Spp4}
\end{equation}
When the surrounding medium is air, the imaginary component of the beam impedance $R$ is small compared with its real part. Therefore, the total impedance magnitude $|r_t|$ can be approximated by the total damping $c_t$ ($|r_t| \approx c_t$). For example, this approximation introduces less than a 4\% deviation for a 2~$\mu$m-wide microcantilever over the frequency range of 20~Hz to 20~kHz. Under this condition, the input-referred thermomechanical noise at each vibration mode (Eq.~\ref{eq:Spp4}) can be approximated as
\begin{equation}
S_{pp_{\text{th},j}}
\approx (\rho_0 c)^2\frac{k_B T }{4 \alpha_j^2 \pi c_t},
\label{eq:Spp4 air}
\end{equation}
which is a constant, independent of frequency, and inversely proportional to the total damping $c_t$. 

Achieving a lower detectable signal is critical for advancing sensor performance. The analytical results highlight two key design factors that enhance the performance of viscous-driven flow sensors, consistent with findings from continuous-system modeling~\cite{Lai2024}. First, the minimum detectable signal level can be improved by increasing the total damping of the mechanical sensing element, a result that contrasts with conventional mechanical sensors such as pressure-based acoustic devices, where higher damping typically degrades performance~\cite{Miles2020}. Second, because each degree of freedom contributes a thermal noise energy of $k_\mathrm{B}T/2$, producing a fixed amount of input-referred noise (Eq.~\ref{eq:Spp4 air}), as is well-known, it is advantageous to design the sensing element with as few vibration modes as possible within the frequency band of interest, ideally a single dominant mode, to minimize input-referred thermomechanical noise~\cite{Miles2020}.

\begin{figure*}[htbp]
  \centering
  \includegraphics[width=0.94\textwidth]{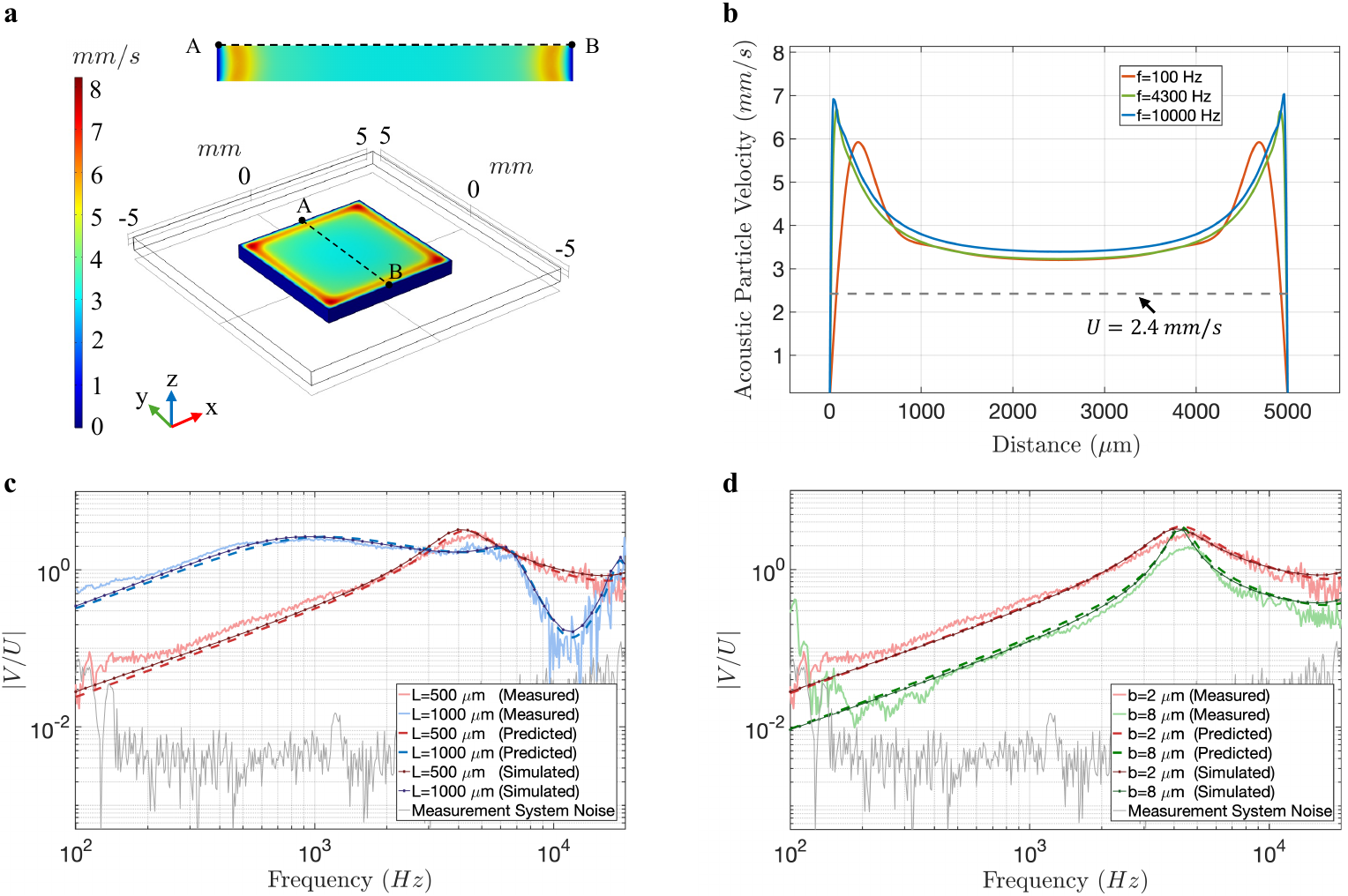} 
\caption{\textbf{Sound-induced velocity responses in slender microcantilevers.} 
Sound waves were generated using loudspeakers placed 3~m from the microcantilever chip and propagated perpendicularly to the chip surface.  
\textbf{a}, Acoustic particle velocity within the chip frame at 100 Hz, obtained from finite element analysis (FEA) under an acoustic pressure of 1~Pa, corresponding to a planar-wave particle velocity of $U = 2.4~\text{mm/s}$. The inset shows the velocity field amplitude across the chip thickness along line A–B.  
\textbf{b}, Velocity amplitude profile along line A–B, as indicated in \textbf{a}.  
\textbf{c}, \textbf{d}, Velocity responses of microcantilevers with varying length ($b = 2~\mu$m) and varying width ($L = 500~\mu$m), respectively. Results from three approaches show excellent agreement: theoretical predictions from Eq.~(\ref{eq:transfer function}) (dashed lines), experimental measurements from the laser Doppler vibrometer (solid lines), and FEA simulations (dotted lines).
}
\label{fig.2} 
\end{figure*}

\subsection{Experimental validation}
In this section, we experimentally validated the theoretical predictions for the velocity response described in Eq.~(\ref{eq:transfer function}), the thermomechanical motion in Eq.~(\ref{eq:Szz_dot}), and the corresponding input-referred noise in Eq.~(\ref{eq:Spp3}) of the microbeam, using the setup illustrated in Fig.~\ref{fig.1}b. To measure the acoustic flow response, loudspeakers were positioned 3~m from the microcantilever chip inside the anechoic chamber to generate a nearly planar acoustic wave over the frequency range of 100~Hz–20~kHz, with sound propagating normal to the chip surface ($\theta=0^{\circ}$). The acoustic particle velocity near the chip was obtained from the measured sound pressure using $U=P/\rho_0c$~\cite{Zhou2017}.

Ideally, in the absence of surrounding boundaries, the acoustic field around a microbeam can be approximated as a uniform plane wave along its length. In practice, however, the chip frame supporting the beam perturbs the local flow, resulting in a nonuniform velocity distribution within the frame opening. To account for this effect, we performed finite element analysis (FEA) to compute the acoustic velocity field inside the chip frame (see Methods). The simulated results, shown in Fig.~\ref{fig.2}a,b, reveal a frequency-dependent, spatially varying velocity field within the chip frame, denoted as $U_{\text{frame}}(x,\omega)$. This effect is incorporated into the experimental validation by updating the mode-dependent coefficient in Eq.~(\ref{eq:feff2}) as $\alpha_j = \frac{1}{L} \int_0^L \phi_j(x)|U_{\text{frame}}(x,\omega)/U|dx$, where $U$ is the amplitude of the incident planar particle velocity. 


Figures~\ref{fig.2}c,d show the sound-induced velocity responses of slender microcantilevers with varying lengths and widths. The dashed lines indicate theoretical predictions, the solid lines represent experimental measurements, and the dotted lines show results from FEA simulations. The strong agreement between the theoretical prediction from Eq.~(\ref{eq:transfer function}) and both the experimental and FEA results across different beam geometries validates the accuracy of the theoretical model. Notably, the 1000~$\mu$m-long microcantilever ($b = 2~\mu$m, $h = 780$~nm) demonstrates high fidelity and large mechanical responsivity over the 100–10,000~Hz range, consistent with the observations from the speech measurement  (Fig.~\ref{fig.1}c). The mechanical responsivity can be further enhanced by reducing the cross-sectional area, allowing viscous forces to dominate more strongly over the mechanical restoring forces, as described in the analytical modeling section.

\begin{figure*}[htbp]
  \centering
  \includegraphics[width=0.94\textwidth]{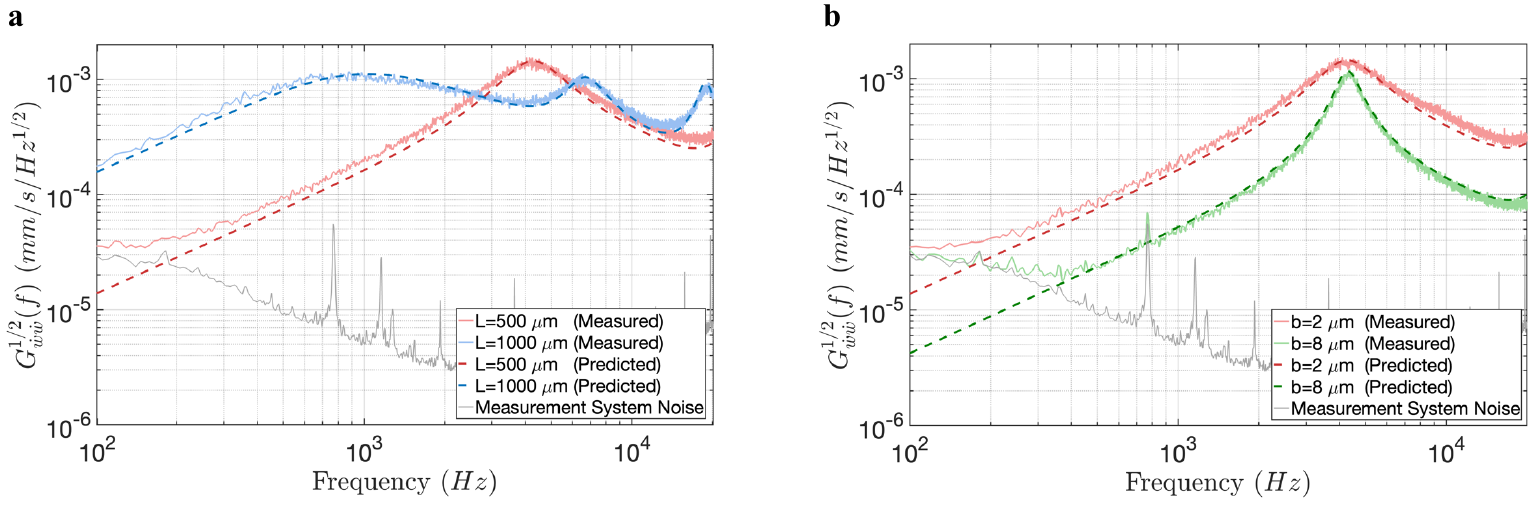} 
\caption{\textbf{Thermomechanical motion in slender microcantilevers.}  
\textbf{a}, \textbf{b}, Square root of the single-sided velocity power spectral density, $G^{1/2}_{\dot{z}\dot{z}}(f)$, for microcantilevers with varying length ($b = 2~\mu\mathrm{m}$) and varying width ($L = 500~\mu\mathrm{m}$), respectively. $G^{1/2}_{\dot{z}\dot{z}}(f)$ is related to the two-sided spectral density by $G_{\dot{z}\dot{z}}(f) = 4\pi S_{\dot{z}\dot{z}}(\omega)$. Dashed lines indicate thermal noise predictions from Eq.~(\ref{eq:Szz_dot}), and solid lines represent experimental data measured using a laser Doppler vibrometer. The experimental system noise floor is shown in gray for reference.
}
\label{fig.3}
\end{figure*}

\begin{figure*}[htbp]
  \centering
  \includegraphics[width=0.94\textwidth]{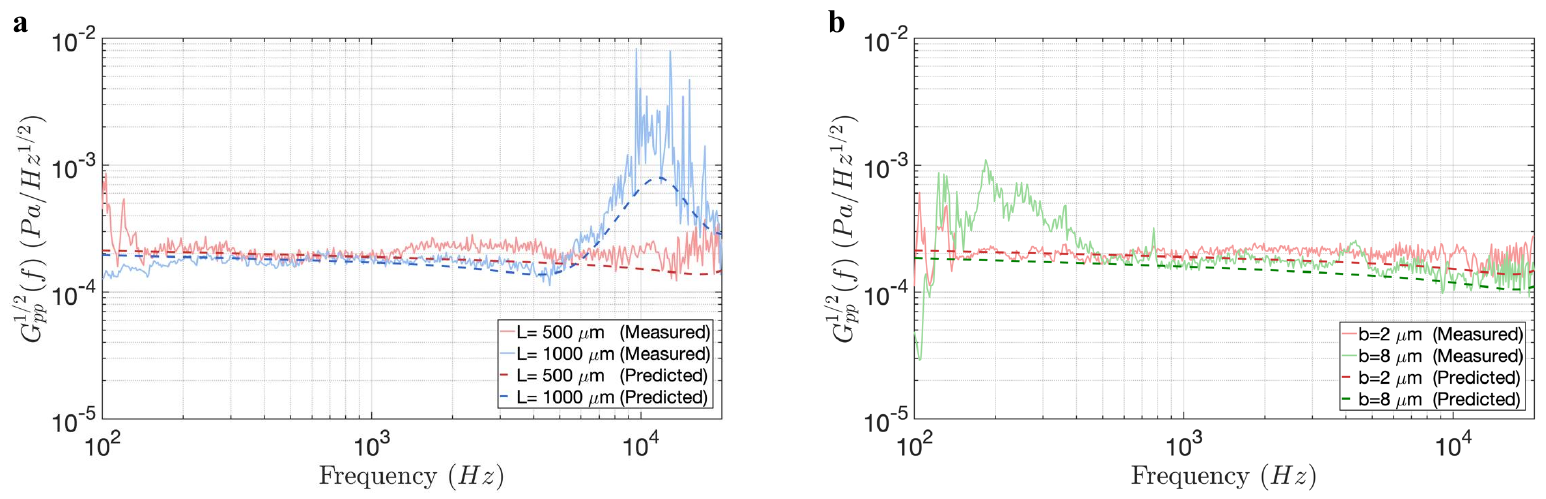} 
\caption{\textbf{Minimum detectable sound limited by thermomechanical noise in slender microcantilevers.}  
\textbf{a}, \textbf{b}, Square root of the single-sided pressure-referred thermomechanical noise, $G^{1/2}_{pp}(f)$, for microcantilevers with varying length ($b = 2~\mu\mathrm{m}$) and varying width ($L = 500~\mu\mathrm{m}$), respectively. Dashed lines indicate theoretical predictions from Eq.~(\ref{eq:Spp3}), and solid lines represent experimental data.
}

\label{fig.4}
\end{figure*}

We further characterized the thermomechanical motion of the slender microcantilevers (Fig.~\ref{fig.3}). The dashed lines in Fig.~\ref{fig.3} indicate the theoretical predictions of thermal noise from Eq.~(\ref{eq:Szz_dot}), the solid lines represent experimental data measured using the laser Doppler vibrometer, and the system noise floor is shown in gray for reference. The experimental data for microcantilevers of varying lengths and widths show excellent agreement with the theoretical predictions across most of the measured frequency range, spanning multiple vibration modes. Deviations below 300~Hz are caused by the measurement system noise.



Based on the measured acoustic response (Fig.~\ref{fig.2}) and thermomechanical motion spectrum (Fig.~\ref{fig.3}), the corresponding pressure-referred thermomechanical noise of the microbeam can be obtained. Figure~\ref{fig.4}a,b present the square root of the single-sided pressure-referred thermomechanical noise $G^{1/2}_{pp}$ for microcantilevers with varying lengths and widths. The experimental data show good agreement with the theoretical predictions from Eq.~(\ref{eq:Spp3}) across most of the frequency range, while deviations at low frequencies are induced by the measurement system noise. Notably, within the measurement frequency range, the 500~$\mu$m-long cantilever primarily exhibits the fundamental flexural mode, resulting in an input-referred noise that remains nearly constant, consistent with the theoretical prediction of Eq.~(\ref{eq:Spp4 air}). In contrast, the 1000~$\mu$m-long cantilever supports multiple vibration modes within the same frequency range, resulting in a prominent peak near 10~kHz.

Collectively, the experimental results  validate the lumped-element model for predicting the flow-induced velocity response, thermomechanical noise, and the minimum detectable signal level, demonstrating strong agreement with measurements across microcantilevers of different dimensions. The simple lumped-element model outperforms the previously developed continuous model in accuracy, particularly at low frequencies~\cite{Lai2024}. The minimum detectable signal can be improved by increasing total damping and minimizing the number of vibration modes within the frequency band of interest, consistent with  continuous-system modeling~\cite{Lai2024}. Interestingly, this design strategy is also evident in mosquitoes, which possess the most sensitive velocity ears known in the animal kingdom~\cite{pantoja2023tiny,menda2019long}. Their three-dimensional, fluffy antennae enhance viscous damping while maintaining a dominant torsional vibration mode within the biologically relevant frequency range for acoustic sensing. This configuration exemplifies how increased total damping can coexist with minimal modal complexity, representing an efficient evolutionary solution for high-sensitivity velocity sensing. In engineered systems, this principle can be implemented through geometric design by arranging multiple slender microstructures in either a two-dimensional configuration (e.g., a perforated microplate~\cite{lin2018nanocardboard}) or a three-dimensional configuration (e.g., a microlattice~\cite{meza2014strong} inspired by mosquito antennae). 

\section{Conclusion}
This study presents a lumped-element model that captures the motion of slender microcantilevers immersed in fluid, combining analytical simplicity with quantitative precision. The model predicts the broadband flow-induced velocity response, thermomechanical noise, and the minimum detectable signal level, showing strong agreement with experimental measurements in air across microcantilevers of varying dimensions. This lumped formulation provides straightforward theoretical guidance for designing high-performance micro- and nanomechanical sensors for flow and vector sound detection.

\section*{Methods}
\textbf{Fabrication}. The fabrication process consists of five main steps. First, a 780~nm-thick $Si_3N_4$ film is deposited on both sides of a double-side-polished silicon wafer (thickness: 525~$\pm$25~$\mu$m) using low-pressure chemical vapor deposition (\textit{Step 1}). The frontside layer serves as the device layer, while the backside layer functions as a mask for subsequent silicon wet etching. The $Si_3N_4$ film on the frontside is patterned by optical lithography and subsequently dry etched with CHF$_3$ (\textit{Step 2}). The backside is then patterned and etched using the same process as the frontside (\textit{Step 3}). After removing the photoresist, the silicon wafer is wet etched in KOH to release the cantilevers (\textit{Step 4}). Finally, the chip is transferred to methanol, which has a relatively low surface tension, then gently fished out and dried with nitrogen gas (\textit{Step 5}). The wet-release and drying process is found to be sufficiently robust to produce the cantilevers investigated in this study. For more delicate structures with larger length-to-thickness ratios, critical-point drying is recommended to minimize stiction and prevent mechanical damage.

\vspace{12pt}
\textbf{Viscous force calculation.} The viscous force
$F$ acting on a beam with a circular cross-section due to the surrounding fluid was first derived by Stokes~\cite{stokes1851internal} and can be written as~\cite{miles2018sound}
\begin{equation}
F(\omega) = R(\omega) V_{\text{r}},
\label{eq:hydrodynamic load}
\end{equation}
where $V_{\text{r}}$ is the relative velocity between the cylinder and the fluid, and $R(\omega)$ represents the impedance of the cylinder per unit length, given by
\begin{equation}
R(\omega) = \frac{\pi}{4} \rho_f \omega r^2 i \Gamma(\omega),
\label{eq:C_f}
\end{equation}
where $r$ is the radius of the cylinder, $\rho_f$ is the density of the surrounding fluid, and $\Gamma(\omega)$ is a frequency-dependent function.
The exact analytical result for $\Gamma(\omega)$ is given by \cite{stokes1851internal}
\begin{equation}
\Gamma_{\text{circ}}(\omega) = 1 + \frac{4\, K_1(\sqrt{i\,\mathrm{Re}})}{\sqrt{i\,\mathrm{Re}}\, K_0(\sqrt{i\,\mathrm{Re}})}
\end{equation}
where Reynolds number $Re=\rho_f \omega b^2/(4 \eta)$ and $\eta$ is the dynamic viscosity; the functions $K_0$ and $K_1$ are modified Bessel functions of the second kind. 

There is no exact analytical expression for $\Gamma(\omega)$ for a beam with a rectangular cross-section. However, an approximate form can be obtained, by introducing a frequency-dependent correction factor $\Omega(\omega)$~\cite{10.1063/1.368002}:
\begin{equation}
\Gamma_{\text{rect}}(\omega) = \Omega(\omega)\, \Gamma_{\text{circ}}(\omega)
\end{equation}
where $\Omega(\omega) = \Omega_r(\omega) + i\Omega_i(\omega)$ is a fitting function that remains accurate to within 0.1\% for Reynolds numbers in the range $Re \in [10^{-6}, 10^{4}]$. The real and imaginary parts are given by
\begin{align}
\Omega_r(\omega) &= 
\big(0.91324 - 0.48274\tau + 0.46842\tau^2 - 0.12886\tau^3 \notag \\
&\quad + 0.044055\tau^4 - 0.0035117\tau^5 + 0.00069085\tau^6\big) \notag \\
&\quad \times \big(1 - 0.56964\tau + 0.48690\tau^2 - 0.13444\tau^3 \notag \\
&\quad + 0.045155\tau^4 - 0.0035862\tau^5 + 0.00069085\tau^6\big)^{-1}
\end{align}
\begin{align}
\Omega_i(\omega) &= 
\big(-0.024134 - 0.029256\tau + 0.016294\tau^2 \notag \\
&\quad - 0.00010961\tau^3 + 0.000064577\tau^4 - 0.00004451\tau^5\big) \notag \\
&\quad \times \big(1 - 0.59702\tau + 0.55182\tau^2 - 0.18357\tau^3 \notag \\
&\quad + 0.079156\tau^4 - 0.014369\tau^5 + 0.0028361\tau^6\big)^{-1}
\end{align}
where $\tau = \log_{10} Re$.

\vspace{12pt}
\textbf{FEA modeling.} Finite element analysis was performed in COMSOL Multiphysics to simulate the dynamic response of a cantilever beam under acoustic excitation in air. The model coupled pressure acoustics, thermoacoustics, and solid mechanics to capture the fluid–structure interactions. The simulation geometry replicated the experimental configuration, consisting of a silicon chip with a central open frame supporting a silicon nitride microcantilever whose dimensions were identical to those used in the experiments.
Material properties of silicon nitride were assigned as follows: Young’s modulus 250~GPa, density 3100~kg/m$^3$. A uniform  pressure of 1~Pa was applied to the chip surface to simulate incident acoustic excitation. The motion at the cantilever tip was extracted to evaluate its acoustic response.

\begin{acknowledgments}
 J.Z. acknowledges startup funding from the State University of New York at Binghamton and research support from NSF Grant No. 2428731. Fabrication was performed at the Center for Nanoscale Materials, a U.S. Department of Energy Office of Science User Facility, and was supported by the U.S. DOE, Office of Basic Energy Sciences, under Contract No. DE-AC02-06CH11357. Film deposition was performed at the Cornell NanoScale Facility, an NNCI member supported by NSF Grant NNCI-2025233.
\end{acknowledgments}

\bibliographystyle{unsrt}
\bibliography{reference}

\begin{thebibliography}{10}

\bibitem{kindig2023asymmetric}
Kayla Kindig, Ruben Stepanyan, Katie~S Kindt, and Brian~M McDermott.
\newblock Asymmetric mechanotransduction by hair cells of the zebrafish lateral line.
\newblock {\em Current Biology}, 33(7):1295--1307, 2023.

\bibitem{coombs2001smart}
Sheryl Coombs.
\newblock Smart skins: information processing by lateral line flow sensors.
\newblock {\em Autonomous robots}, 11(3):255--261, 2001.

\bibitem{menda2019long}
Gil Menda, Eyal~I Nitzany, Paul~S Shamble, Amelia Wells, Laura~C Harrington, Ronald~N Miles, and Ronald~R Hoy.
\newblock The long and short of hearing in the mosquito aedes aegypti.
\newblock {\em Current Biology}, 29(4):709--714, 2019.

\bibitem{pantoja2023tiny}
Hoover Pantoja-S{\'a}nchez, Brian~C Leavell, Bianca Rendon, WA~Priyanka~P de~Silva, Richa Singh, Jian Zhou, Gil Menda, Ronald~R Hoy, Ronald~N Miles, Neil~D Sanscrainte, et~al.
\newblock Tiny spies: mosquito antennae are sensitive sensors for eavesdropping on frog calls.
\newblock {\em Journal of Experimental Biology}, 226(24):jeb245359, 2023.

\bibitem{nakata2020aerodynamic}
Toshiyuki Nakata, Nathan Phillips, Patr{\'\i}cio Sim{\~o}es, Ian~J Russell, Jorn~A Cheney, Simon~M Walker, and Richard~J Bomphrey.
\newblock Aerodynamic imaging by mosquitoes inspires a surface detector for autonomous flying vehicles.
\newblock {\em Science}, 368(6491):634--637, 2020.

\bibitem{gopfert2000nanometre}
Martin~C G{\"o}pfert and Daniel Robert.
\newblock Nanometre--range acoustic sensitivity in male and female mosquitoes.
\newblock {\em Proceedings of the Royal Society of London. Series B: Biological Sciences}, 267(1442):453--457, 2000.

\bibitem{shamble2016airborne}
Paul~S Shamble, Gil Menda, James~R Golden, Eyal~I Nitzany, Katherine Walden, Tsevi Beatus, Damian~O Elias, Itai Cohen, Ronald~N Miles, and Ronald~R Hoy.
\newblock Airborne acoustic perception by a jumping spider.
\newblock {\em Current Biology}, 26(21):2913--2920, 2016.

\bibitem{zhou2022}
Jian Zhou, Junpeng Lai, Gil Menda, Jay~A. Stafstrom, Carol~I. Miles, Ronald~R. Hoy, and Ronald~N. Miles.
\newblock Outsourced hearing in an orb-weaving spider that uses its web as an auditory sensor.
\newblock {\em Proceedings of the National Academy of Sciences}, 119(14):e2122789119, 2022.

\bibitem{Zhou2017}
Jian Zhou and Ronald~N. Miles.
\newblock Sensing fluctuating airflow with spider silk.
\newblock {\em Proceedings of the National Academy of Sciences}, 114(46):12120--12125, 2017.

\bibitem{Gopfert2000}
Martin~C. Göpfert and Daniel Robert.
\newblock Nanometre-range acoustic sensitivity in male and female mosquitoes.
\newblock {\em Proceedings of the Royal Society of London. Series B: Biological Sciences}, 267:453--457, 2000.

\bibitem{tao2012hair}
Junliang Tao and Xiong~Bill Yu.
\newblock Hair flow sensors: from bio-inspiration to bio-mimicking—a review.
\newblock {\em Smart Materials and Structures}, 21(11):113001, 2012.

\bibitem{droogendijk2014performance}
Harmen Droogendijk, J{\'e}rome Casas, Thomas Steinmann, and GJM Krijnen.
\newblock Performance assessment of bio-inspired systems: flow sensing mems hairs.
\newblock {\em Bioinspiration \& biomimetics}, 10(1):016001, 2014.

\bibitem{kottapalli2016nanofibril}
Ajay Giri~Prakash Kottapalli, Meghali Bora, Mohsen Asadnia, Jianmin Miao, Subbu~S Venkatraman, and Michael Triantafyllou.
\newblock Nanofibril scaffold assisted mems artificial hydrogel neuromasts for enhanced sensitivity flow sensing.
\newblock {\em Scientific reports}, 6(1):19336, 2016.

\bibitem{jin2025flexible}
Biao Jin, Hongchao Cao, Tianyu Sheng, Zheng Gong, Zihao Dong, Yansong Gai, and Yonggang Jiang.
\newblock Flexible hair-like piezoelectric acoustic particle velocity sensor with enhanced sensitivity for speaker recognition.
\newblock {\em Advanced Functional Materials}, 35(11):2417164, 2025.

\bibitem{hu2025high}
Jinxin Hu, Bin Liu, Jiyuan Feng, et~al.
\newblock High sensitivity acoustic flow sensing based on bio-inspired web-like structure for panoramic acoustic perception.
\newblock {\em Microsystems \& Nanoengineering}, 11:136, 2025.

\bibitem{li2025fish}
Liangye Li, Xuhao Fan, Geng Chen, Yueqi Liu, Fujun Zhang, Zhuolin Chen, Zhi Zhang, Wangyang Xu, Shixiong Zhang, Yuncheng Liu, et~al.
\newblock From fish to fiber: 3d-nanoprinted optical neuromast for multi-integrated underwater detection.
\newblock {\em Nature Communications}, 16(1):7390, 2025.

\bibitem{zhou2018highly}
Jian Zhou, Boxiao Li, Jian Liu, Wayne~E Jones, and Ronald~N Miles.
\newblock Highly-damped nanofiber mesh for ultrasensitive broadband acoustic flow detection.
\newblock {\em Journal of Micromechanics and Microengineering}, 28(9):095003, 2018.

\bibitem{shimozawa1984aerodynamics}
Tateo Shimozawa and Masamichi Kanou.
\newblock The aerodynamics and sensory physiology of range fractionation in the cercal filiform sensilla of the cricket gryllus bimaculatus.
\newblock 1984.

\bibitem{miles2018sound}
RN~Miles and J~Zhou.
\newblock Sound-induced motion of a nanoscale fiber.
\newblock {\em Journal of Vibration and Acoustics}, 140(1):011009, 2018.

\bibitem{Lai2024}
Junpeng Lai, Mahdi Farahikia, Morteza Karimi, Zihan Liu, Yingchun Jiang, Changhong Ke, and Ronald Miles.
\newblock Effect of size on the thermal noise and acoustic response of viscous-driven microbeams.
\newblock {\em The Journal of the Acoustical Society of America}, 155(4):2561--2576, 04 2024.

\bibitem{doi:https://doi.org/10.1002/9780470117866.ch11}
Singiresu~S. Rao.
\newblock {\em Transverse Vibration of Beams}, chapter~11, pages 317--392.
\newblock John Wiley \& Sons, Ltd, 2006.

\bibitem{Miles2020}
Ronald~N. Miles.
\newblock {\em Physical Approach to Engineering Acoustics}.
\newblock Springer Cham, 2020.

\bibitem{lin2018nanocardboard}
Chen Lin, Samuel~M Nicaise, Drew~E Lilley, Joan Cortes, Pengcheng Jiao, Jaspreet Singh, Mohsen Azadi, Gerald~G Lopez, Meredith Metzler, Prashant~K Purohit, et~al.
\newblock Nanocardboard as a nanoscale analog of hollow sandwich plates.
\newblock {\em Nature communications}, 9(1):4442, 2018.

\bibitem{meza2014strong}
Lucas~R Meza, Satyajit Das, and Julia~R Greer.
\newblock Strong, lightweight, and recoverable three-dimensional ceramic nanolattices.
\newblock {\em Science}, 345(6202):1322--1326, 2014.

\bibitem{stokes1851internal}
George~Gabriel Stokes.
\newblock On the effect of the internal friction of fluids on the motion of pendulums.
\newblock {\em Transactions of the Cambridge Philosophical Society}, 9:8--106, 1851.

\bibitem{10.1063/1.368002}
John~Elie Sader.
\newblock Frequency response of cantilever beams immersed in viscous fluids with applications to the atomic force microscope.
\newblock {\em Journal of Applied Physics}, 84(1):64--76, 07 1998.

\end{thebibliography}

\end{document}